\documentclass[12pt]{iopart}
\usepackage{iopams}
\usepackage{amssymb}
\usepackage{amscd}
\usepackage{wasysym}
\usepackage[ansinew]{inputenc}
\usepackage[T1]{fontenc}
\usepackage{ae,aecompl}
\usepackage{cite}

\usepackage[dvips]{graphicx}
\DeclareGraphicsExtensions{.eps}

\newcommand{\be}{\begin{equation}}
\newcommand{\ee}{\end{equation}}
\newcommand{\bea}{\begin{eqnarray}}
\newcommand{\eea}{\end{eqnarray}}
\newcommand{\nn}{\nonumber}

\renewcommand{\vec}[1]{\mathbf{#1}}

\begin{document}

\title[Mean-field dynamics of a two-mode BEC subject to noise and dissipation]{Mean-field dynamics of a two-mode Bose-Einstein condensate subject to noise and dissipation}
\author{F Trimborn$^1$, D Witthaut$^2$ and S Wimberger$^3$}
\address{${}^1$ Institut f\"ur mathematische Physik, TU Braunschweig, D--38106 Braunschweig, Germany \\
				 ${}^2$ QUANTOP, Niels Bohr Institute, University of Copenhagen, DK--2100 Copenhagen, Denmark \\
				 ${}^3$ Institut f\"ur theoretische Physik, Universit\"at Heidelberg, D--69120, Heidelberg, Germany}
\ead{dirk.witthaut@nbi.dk}
\date{\today }

\begin{abstract}
We discuss the dynamics of an open two-mode Bose-Hubbard system subject to phase noise 
and particle dissipation. Starting from the full many-body dynamics 
described by a master equation the mean-field limit is derived resulting in
an effective non-hermitian (discrete) Gross-Pitaevskii equation which has been 
introduced only phenomenologically up to now.
The familiar mean-field phase space structure is substantially altered by the 
dissipation. Especially the character of the fixed points shows an abrupt 
transition from elliptic or hyperbolic to attractiv or repulsive, respectively.
This reflects the metastable behaviour of the corresponding many-body system which 
surprisingly also leads to a significant increase of the purity of the condensate. 
A comparison of the mean-field approximation to simulations of the full master equation using 
the Monte Carlo wave function method shows an excellent agreement for wide 
parameter ranges.
\end{abstract}

\pacs{03.75.Lm, 03.75.Gg, 03.65.Yz}
\maketitle


The physics of ultracold atoms in optical lattices has made an enormous 
progress in the last decade, as it is an excellent model system 
for a variety of fields such as nonlinear dynamics or condensed matter 
physics \cite{Mors06,Madr06}. Although this seems to be an inherent many-particle problem, 
the dynamics of the macroscopic wave function is remarkably well 
reproduced by the (discrete) Gross-Pitaevskii equation (GPE) if the 
systems undergoes a Bose-Einstein condensation \cite{Pita03}. 
Recently there has been an increased theoretical \cite{Angl97,Ruos98,Angl01,Pono06,Wang07} 
as well as experimental \cite{Syas08} interest in the dynamics of these 
systems coupled to the environment. In particular, the effects of particle 
loss have been discussed in several theoretical articles resorting to an effective 
non-hermitian mean-field description introduced phenomenologically 
to analyse resonances, transport and localization effects 
\cite{Mois05,Schl06,Witt06,Livi06,Hill06,Schl07,Grae08}.

In this communication we want to illuminate the origin and give a convincing 
motivation of this approach. Starting from a master equation describing the full 
many-body dynamics including phase noise and particle loss we derive a 
generalized, non-hermitian Gross-Pitaevskii equation. Due to the decay the structure 
of the resulting dynamics abruptly changes introducing repulsive and attractive 
fixed points. Unlike the dissipation-free case there are no longer oscillations around 
the fixed points, such that irreversible transitions between the former self-trapping fixed points 
are possible. This reflects the metastable behaviour of the open many-particle system
and gives rise to a significant purification of the BEC by the dissipation which will be 
explained here. The validity of the 
presented approximation is tested by a comparison to full quantum many-body 
calculations, showing that the mean-field approximation provides an excellent 
description of the system. To integrate the effective description by a non-hermitian 
GPE into well-known concepts from the theory of open quantum systems, we discuss their 
relation to the quantum jump approach \cite{Dali92,Carm93}, demonstrating that they provide a well-suited tool 
to analyse the short- as well as the long-time behaviour of the open many-particle system.    

In particular, we consider the dynamics of ultracold atoms in an open double-well 
trap which is not only an extremely popular model system, but also has several 
recent experimental realisations \cite{Albi05,Gati06,Schu05b,Foll07}. 
The unitary part of the dynamics is described by the Bose-Hubbard type Hamiltonian 
\bea
  \hat H &=& -J  \left( \hat a_1^\dagger \hat a_2 +  \hat a_2^\dagger \hat a_1 \right)
     + \epsilon_1 \hat n_1 +  \epsilon_2 \hat n_2  + \frac{U}{2} \left( \hat a_1^{\dagger 2} \hat a_1^2 
            + \hat a_2^{\dagger 2} \hat a_2^2  \right),
    \label{eqn-hami-bh}
\eea
where $\hat a_j$ and $\hat a_j^\dagger$ are the bosonic annihilation and creation
operators in mode $j$ and $\hat n_j = \hat a_j^\dagger \hat a_j$
is the correponding number operator. We set $\hbar = 1$, thus measuring 
all energies in frequency units.
In order to analyze the dynamics in the Bloch--representation we transform the 
Hamiltonian using the collective operators
\bea
  &&\hat L_x = \frac{1}{2} 
      \left( \hat a_1^\dagger \hat a_2  + \hat a_2^\dagger \hat a_1 \right), \quad 
  \hat L_y = \frac{\rmi}{2} 
      \left( \hat a_1^\dagger \hat a_2  - \hat a_2^\dagger \hat a_1 \right),
      \nn \\
  && \qquad \qquad \hat L_z = \frac{1}{2} 
      \left( \hat a_2^\dagger \hat a_2  - \hat a_1^\dagger \hat a_1 \right),
  \label{eqn-angular-op} 
\eea
which form an angular momentum algebra $su(2)$ with rotational quantum number $N/2$
\cite{Milb97,Smer97,Angl01,08phase}.
With these definitions the Hamiltonian (\ref{eqn-hami-bh}) can be rewritten as
\be
  \hat H =  -2 J \hat L_x + 2 \epsilon \hat L_z   + U \hat L_z^2
  \label{eqn-hamiltonian-2level}
\ee
with $2\epsilon = \epsilon_1 - \epsilon_2$ up to terms which only depend 
on the total number of atoms. The macroscopic dynamics of the atomic 
cloud is well described within a mean-field approximation, only considering 
the expectation values of the angular momentum operators 
$\ell_j = \langle \hat L_j \rangle$ and the particle number 
$n = \langle \hat n_1 + \hat n_2 \rangle$ \cite{Milb97,Smer97,Angl01}. 

Here we consider the dissipative extension of this system. 
The main source of decoherence in current experiments is phase noise due 
to elastic collision with atoms in the thermal cloud \cite{Angl97,Ruos98} 
which effectively heats the system, leaving the particle number invariant.
Only recently, methods to tame this source of decoherence were discussed 
in \cite{Khod08,08stores}. In this communication 
we focus on the effects of particle loss at constant rates $\gamma_{aj}$ 
in the two wells $j=1,2$. 
Such a loss is not only of fundamental interest but can be adapted in current
experiments without greater difficulties by removing atoms with a focused resonant
laser beam or by inducing a radio frequency transition to an untrapped 
internal state \cite{Bloc99}. All parameters used here lie in realistic ranges
for ongoing experiments \cite{Albi05,Gati06}. 

The master equation description including both, phase noise and particle loss, is
well established \cite{Gard04} and routinely used in the context of
photon fields. Thus we consider the dynamics generated by the master 
equation
\bea
  \dot{\hat \rho} &=& -\rmi [\hat H,\hat \rho] 
  - \frac{\gamma_p}{2} \sum_{j = 1,2}
    \left(\hat n_j^2 \hat \rho + \hat \rho \hat n_j^2 - 2 \hat n_j \hat \rho \hat n_j  \right)\nn \\
   && \quad - \frac{1}{2}  \sum_{j=1,2}  \gamma_{aj} \left(
     \hat a_j^\dagger \hat a_j \hat \rho + \hat \rho \hat a_j^\dagger \hat a_j 
    - 2 \hat a_j \hat \rho \hat a_j^\dagger \right).
   \label{eqn-master2}
\eea
The evolution equations for the expectation values of the angular momentum
operators (\ref{eqn-angular-op}) can be calculated starting
from the Master equation via $\dot \ell_j = \tr(\hat L_j \dot {\hat \rho})$ with $j=x,y,z$. 
This yields the exact result
\bea
  \dot \ell_x &=& -2 \epsilon \ell_y - 2U (\ell_y \ell_z + \Delta_{yz}) 
        - T_2^{-1} \ell_x, \nn \\
  \dot \ell_y &=&  2 J \ell_z + 2 \epsilon \ell_x 
        + 2U (\ell_x \ell_z + J_{xz}) - T_2^{-1} \ell_y, \nn \\
  \dot \ell_z &=& - 2 J \ell_y - T_1^{-1} \ell_z - T_1^{-1} f_a n/2, \nn \\
  \dot n   &=& - T_1^{-1} n  - 2 T_1^{-1} f_a \ell_z,
  \label{eqn-eom-bloch1}
\eea
where we have defined the transversal $T_1^{-1}$ and longitudinal $T_2^{-1}$ damping rates by
\be
  T_1^{-1} = (\gamma_{a1} + \gamma_{a2})/2 \quad \mbox{and} \quad
  T_2^{-1} = \gamma_p +  T_1^{-1}
  \label{eqn-rel-times}
\ee
and the asymmetry of the loss rates by
$f_a = (\gamma_{a2}$ -- $\gamma_{a1})/ (\gamma_{a1}$ + $\gamma_{a2})$.
In the non-interacting case $U=0$, these equations of motion ressemble
the Bloch equations in nuclear magnetic resonance \cite{Bloc46}, 
except for the fact that the 'equilibrium' value of the population imbalance 
$\ell_z$ is given by $-f_a n/2$ and therefore depends on the decreasing 
expectation value of the total particle number $n$.

The exact equations of motion (\ref{eqn-eom-bloch1}) still include the covariances
$
  \Delta_{jk} = \langle \hat L_j \hat L_k + \hat L_k \hat L_j  \rangle/2 
   - \langle \hat L_j \rangle \langle \hat L_k \rangle.
$
The approximation of second order moments by products of expectation values, 
such that $\Delta_{jk}\approx0$ yields the well-known mean-field description.
This truncation is valid in the macroscopic limit, since the covarinces
vanish as $1/n$ if the many-particle quantum state is close to a pure BEC.
Here and in the following we depict the rescaled variables 
$s_j = \ell_j/n$, thus renormalizing to separate the decay of the particle number $n$ 
from the internal dynamics.

\begin{figure}[tb]
\centering
\includegraphics[width=8cm, angle=0]{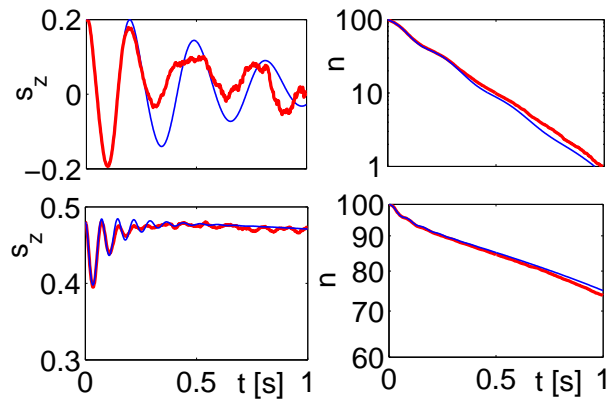}
\caption{\label{fig-mpmf-compare}
Comparison of the mean-field approximation (thin blue line)
with the full many-particle dynamics calculated with the MCWF method (thick 
red line) for $J=10 \, {\rm s}^{-1}$, $U=1 \,{\rm s}^{-1}$, 
$\gamma_p = 3 \,{\rm s}^{-1}$and $\gamma_{a2} = 5 \, {\rm s}^{-1}$.
The initial state was assumed to be a pure BEC (i.e. a product state)
with $\vec s= (0.46,0,0.2) $ (a) and $\vec s= (0.14,0,0.48) $ (b), 
respectively, and $n=100$ particles.
}
\end{figure}

The benefit of the mean-field approximation is illustrated in figure \ref{fig-mpmf-compare}, where it is compared to the full many-particle quantum dynamics calculated with the Monte Carlo wave function (MCWF) method \cite{Dali92,Carm93}.
The trajectory in figure \ref{fig-mpmf-compare} (a) was launched at $\vec s = (0.46,0,0.2)$ with a moderate population imbalance, thus performing Josephson oscillations \cite{Albi05}. The amplitude is damped because of the phase noise, while the oscillation period increases as the effective macroscopic interaction strength $g(t):=Un(t)$ decreases. The 
decay of the particle number $n(t)$ is also strongly modulated by the oscillations of the population imbalance. 
The trajectory in figure \ref{fig-mpmf-compare} (b) was launched at $\vec s = (0.14,0,0.48)$ in the self-trapping region. The residual oscillations are rapidly damped out and the system relaxes to a quasi-steady state on the shown time scale. The particle number decreases slowly and non-exponentially, since the condensate is mostly localized in the non-decaying potential well, cf.~also \cite{Schl07}.
All these features of the dynamics are well reproduced by the mean-field description, and the decay of the particle number is accurately predicted. Strong deviations are only expected in the vicinity of unstable fixed points of the mean-field dynamics, which can be nearly cured within the framework of phase space distributions \cite{08phase}. 

In order to explore the genuine effects of particle loss, phase noise is neglected ($\gamma_p = 0$)
in the following. In this case the dynamics can be further simplified and one can easily show that the 
particle number coincides with the magnitude of the Bloch vector $\sqrt{\ell_x^2+\ell_y^2+\ell_z^2} = n/2$, which we can 
use to reformulate 
the mean-field dynamics by an effective non-hermitian Gross-Pitaevskii equation
\be
  \rmi \frac{\rmd}{\rmd t} \left(\begin{array}{c} \psi_1 \\ \psi_2 \end{array}\right) =
  \left(\begin{array}{cc} \tilde \epsilon_1 + U |\psi_1|^2 & -J \\
  -J & \tilde \epsilon_2  + U |\psi_2|^2 \end{array}\right)
  \left(\begin{array}{c} \psi_1 \\ \psi_2 \end{array}\right)
\ee
with complex on-site energies $\tilde \epsilon_j = \epsilon_j - i \gamma_{aj}/2$.
The equivalence to the Bloch vector description is established via the identification
\bea
   && \ell_x = \frac{1}{2}(\psi_1^* \psi_2 + \psi_2^* \psi_1), \quad 
   \ell_y = \frac{1}{2\rmi}(\psi_2^* \psi_1 - \psi_1^* \psi_2), \nn \\
   && \qquad \qquad \ell_z = \frac{1}{2}(|\psi_2|^2 - |\psi_1|^2)
\eea
and $n = |\psi_1|^2 + |\psi_2|^2$.
In this effective description a loss of particles is represented by a
loss of normalization.

We now consider the dynamics for a fixed value of the macroscopic interaction strength 
$g=Un=const.$ in the special case $\gamma_{a1} =:\gamma$ 
and $\gamma_{a2}=0$. Even though the restriction
to a fixed interaction constant seems to be artificial, it reveals 
the effects of the particle loss on the structure of the mean-field 
phase space and especially the character of the fixed points most clearly. 
Moreover, the dynamics under a fixed interaction constant correspond to the 
periods of constant particle number between two loss processes in the quantum 
jumps picture \cite{Dali92,Carm93}. Therefore this treatment provides a well-suited description 
of the short- as well as the long-time behaviour.
Note that the more general case $\gamma_{a1}\neq0$ and $\gamma_{a2}\neq0$ does not 
lead to a fundamentally different dynamical behaviour since only the difference of 
the decay rates influences the internal dynamics. However, the expectation 
value of the particle number $n$ and thereby also the effective interaction 
strength $g(t)$ decrease faster.

\begin{figure}[tb]
\centering
\includegraphics[width=4cm, angle=0]{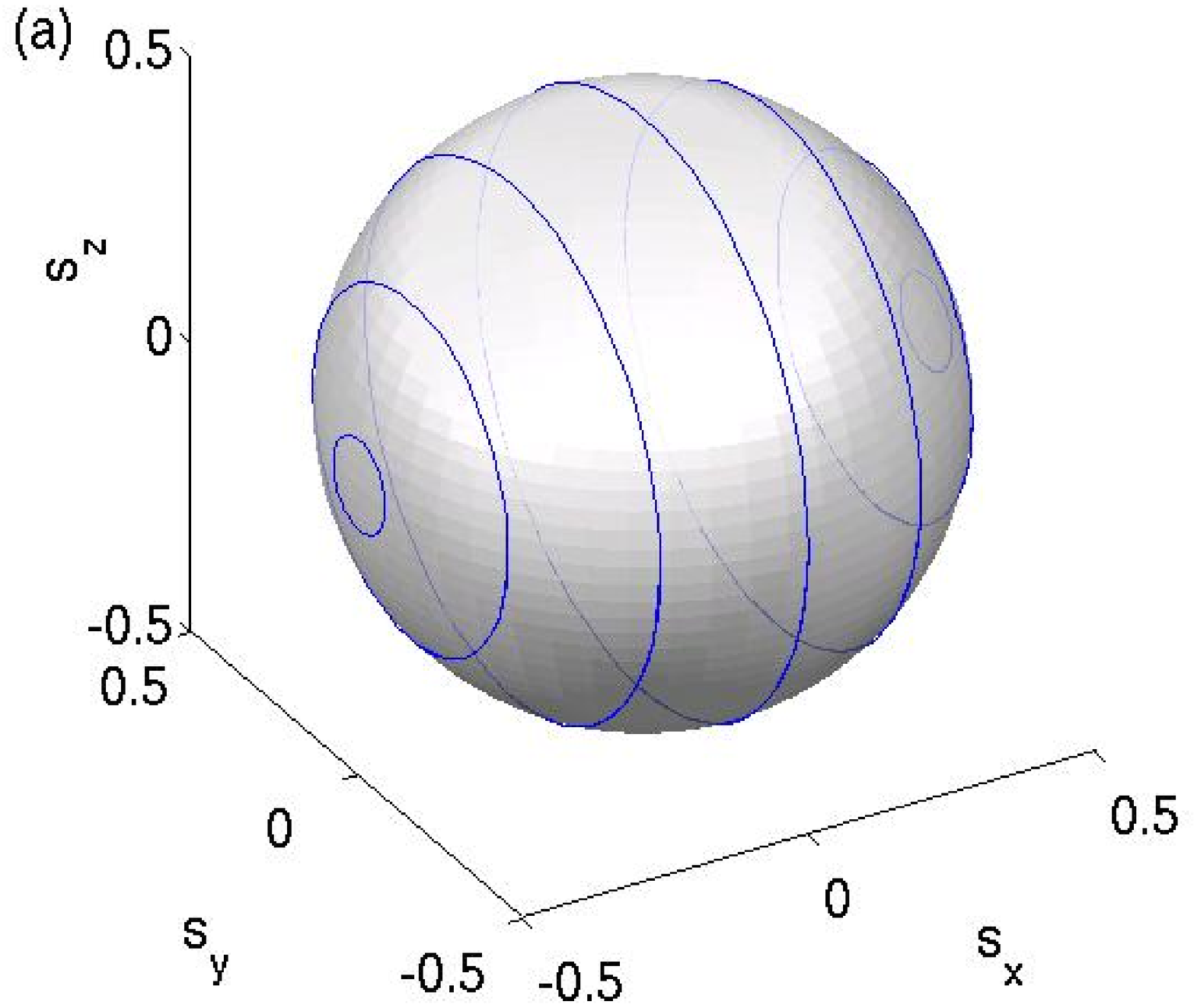}
\includegraphics[width=4cm, angle=0]{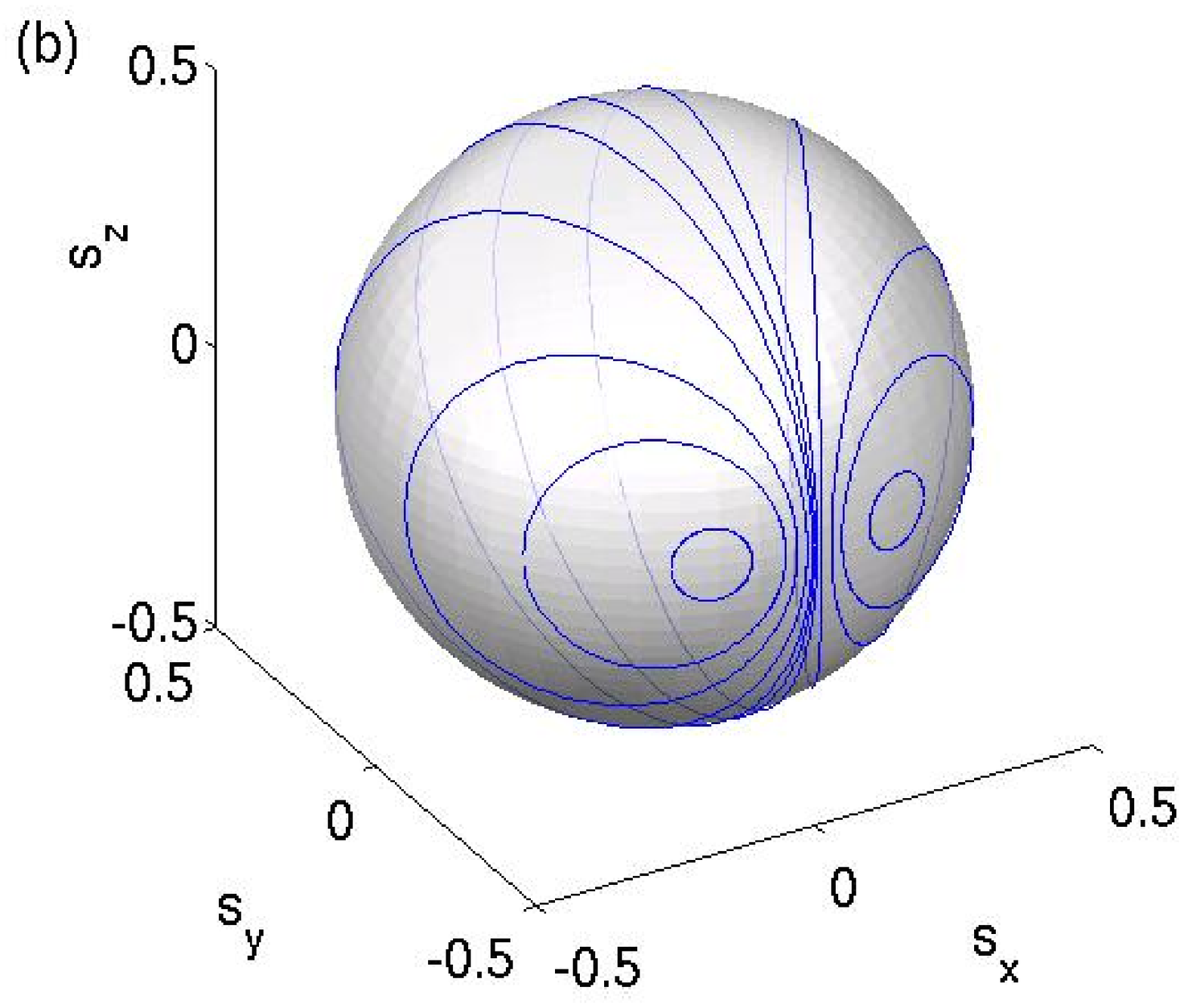}
\includegraphics[width=4cm, angle=0]{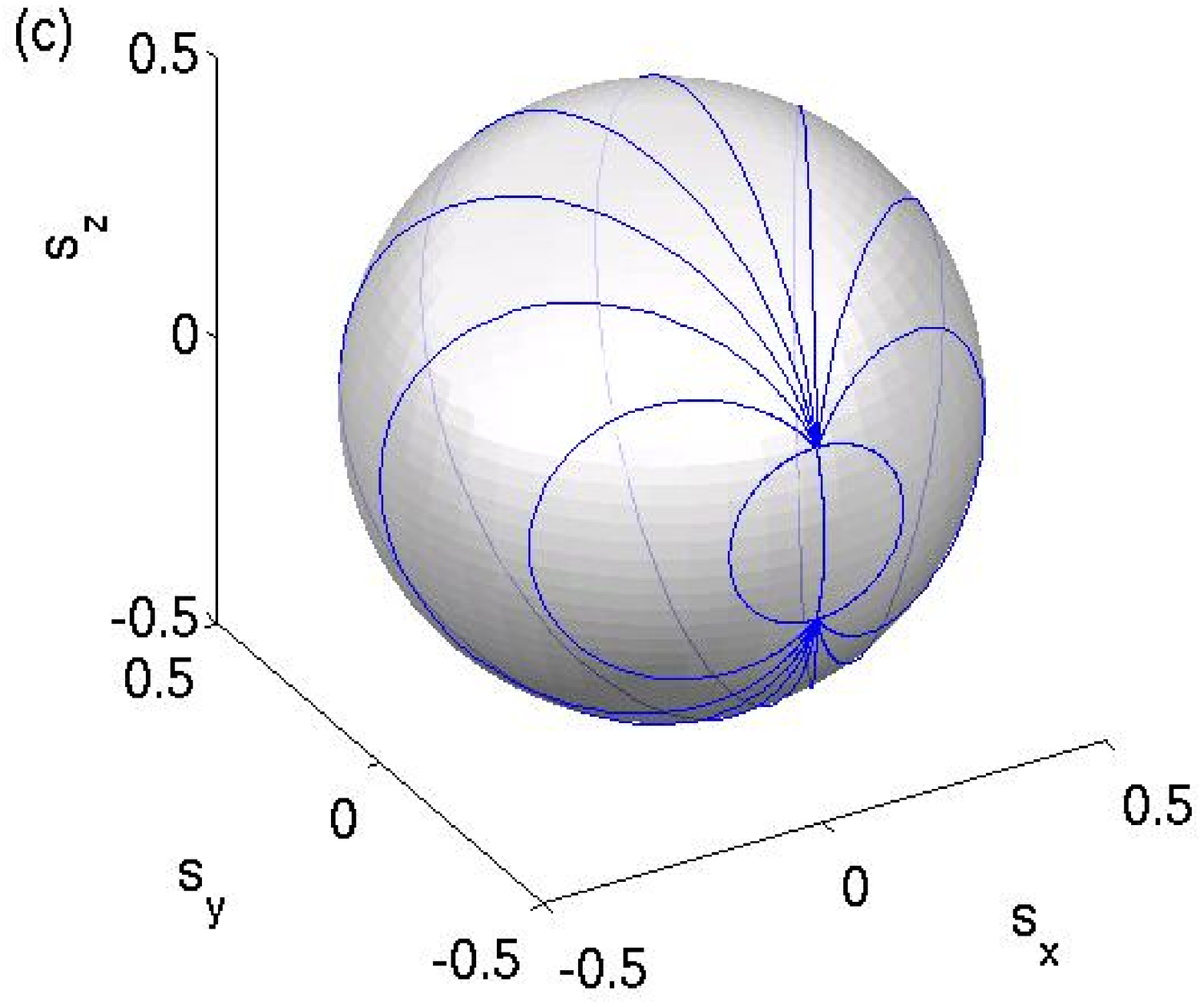}
\includegraphics[width=4cm, angle=0]{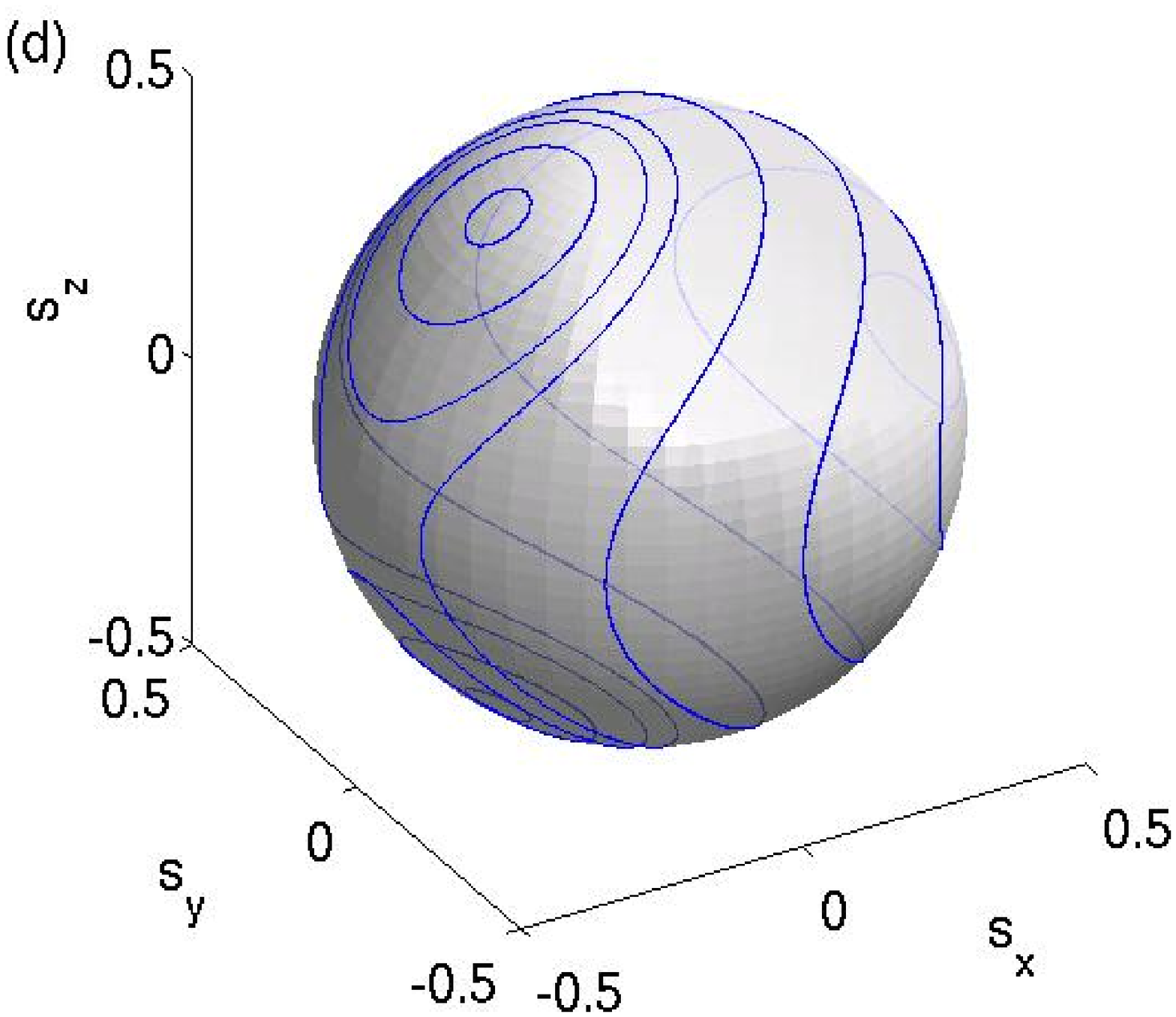}
\includegraphics[width=4cm, angle=0]{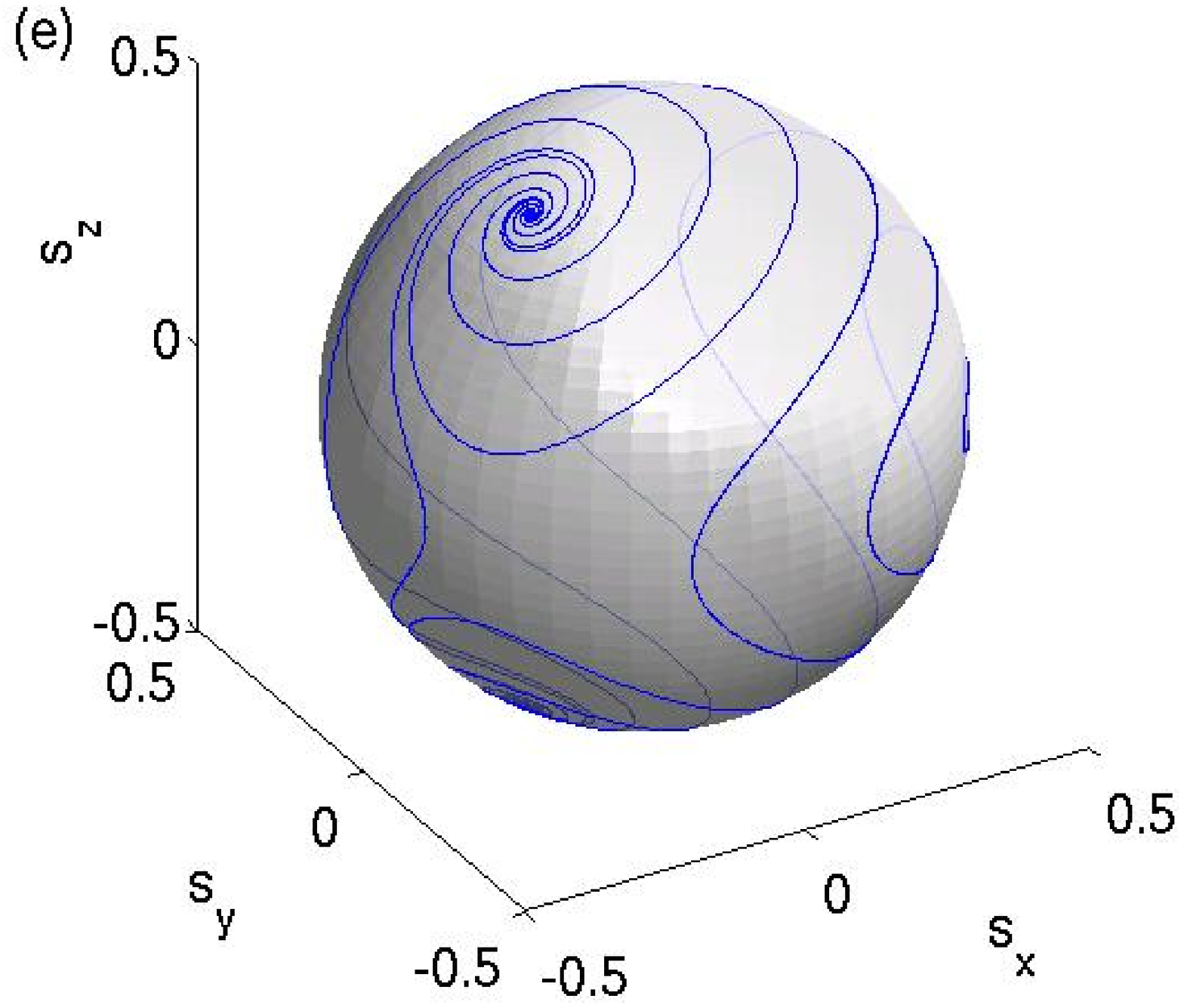}
\includegraphics[width=4cm, angle=0]{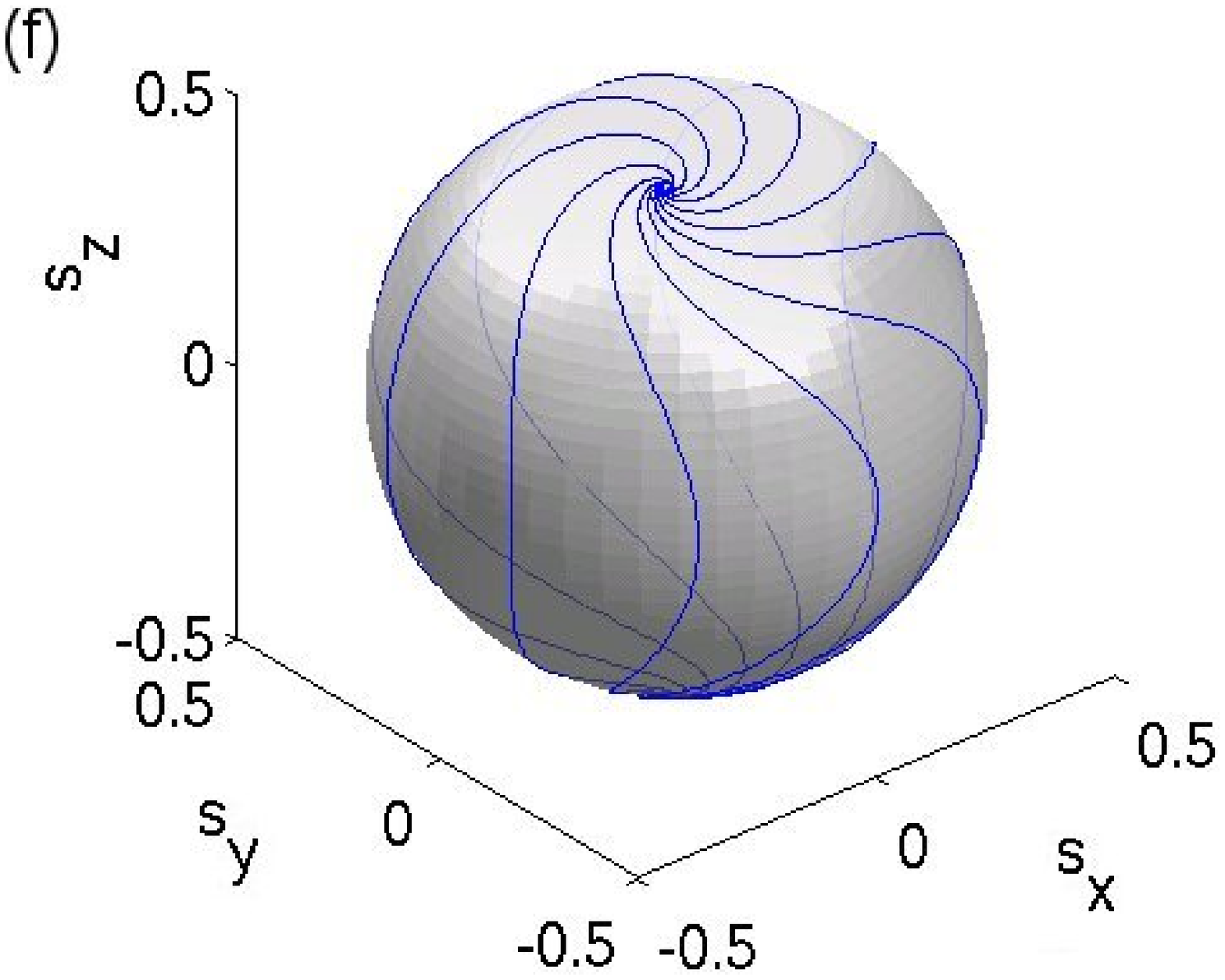}
\caption{\label{fig-withoutU}
Mean-field dynamics for the non-interacting case $g=0$ (upper row) 
and for a fixed interaction strength $g=4 \,{\rm s}^{-1}$ (lower row) 
in dependence of the decay rate ($\gamma=0$ for (a) and (d), 
$\gamma=1.9 \, {\rm s}^{-1}$  for (b), $\gamma=2.1 \, {\rm s}^{-1}$ for (c), 
$\gamma=1 \, {\rm s}^{-1} $ for (e) and $\gamma=4 \, {\rm s}^{-1}$ for (f)) 
-- for all figures holds $J=1 \, {\rm s}^{-1}$ and $\epsilon =0$.
}
\end{figure}

The resulting dynamics of the Bloch vector is illustrated in
figure \ref{fig-withoutU}. The upper row (a-c) shows the phase space 
for the linear case, $U=0$, where the 
mean-field approximation is exact. Without loss one recovers the famous
Josephson oscillations (a). An analysis of the fixed points for the 
dissipative dynamics shows the emergence of two regimes depending on 
the amplitude of the loss rate.
For weak losses, $|\gamma| \leq 2 J$, the fixed points are given by
\be \label{eqn_U0gamma}
\vec s_\pm^J =  \left( \begin{array}{c}
   \pm \left[\frac 1 4 - (\frac{\gamma}{4 J})^2 \right]^{\frac 1 2} \\ 
   - \frac \gamma {4 J} \\ 
   0 \end{array} \right).
\ee  
While the fixed points remain elliptic and the population is still equally 
distributed, the fixed points are no longer symmetric, since the relative 
phase between them decreases (b).
This behaviour can be qualitatively understood within the analogy to Josephson 
junctions: The weak decay induces an assymetry between the wells leading to to 
a continuous particle stream to the first well. At the fixed points this effect
is compensated by the Josephson current $I_J\propto J s_y$ requiring $s_y \neq 0$.

For stronger decay rates, $|\gamma| \geq 2 J$, the two fixed points are given by
\be 
\vec s_{\pm}^D =  \left( \begin{array}{c}
   0 \\ 
   -\frac J \gamma \\ 
   \pm \left[\frac 1 4 - (\frac{J^2}{\gamma^2}) \right]^{\frac 1 2} 
   \end{array} \right).
\ee
Above the critical value $|\gamma| = 2 J$ the character of the two fixed points 
changes abruptly from elliptic into an attractive and a repulsive one as shown in
figure \ref{fig-withoutU} (c). 
The maximal Josephson current is no longer sufficient to compensate the current 
induced by the decay leading to a population excess in the non-decaying 
site. This explains the population imbalance in the fixed points which increases 
with ascending decay rates.

In the strongly interacting case without dissipation one observes the splitting of one of the
elliptic fixed points into two novel elliptic and one hyperbolic fixed point -- this is
the famous self-trapping effect \cite{Milb97,Smer97,Albi05}.
The critical interaction strength for the occurence of this bifurcation is lowered in the presence 
of dissipation to $g^2=U^2n^2\geq 4 J^2-\gamma^2$.
In the subcritical regime for $\gamma < 2 J$ and $Un\leq 4 J^2-\gamma^2$,
we find oscillations around the same fixed points $s_\pm^J$ as in the non-interacting, but 
dissipative case (\ref{eqn_U0gamma}). However, these are now distorted (not shown in the figure).
In the overcritical regime $g=Un > \sqrt{4J^2 + \gamma^2}$ and for a
weak decay $\gamma < 2 J$ one rediscovers a generalized self-trapping effect. As a result of the 
dissipative process, one elliptic fixed point now bifurcates into an attractive and a 
repulsive fixed point (in contrast to the two elliptic ones) and one hyperbolic one (cf. figure \ref{fig-withoutU} (e)). The novel fixed points are located at:
\be
\vec s_\pm^{\pi} =  \frac 1 {\gamma^2 + g^2}\left( \begin{array}{c}
   - g J \\ 
   - \gamma J \\ 
   \pm \sqrt{(\gamma^2 + g^2)(\frac{\gamma^2 + g^2}{4} - J ^2)} 
   \end{array} \right).
\ee
For stronger decay rates, $\gamma \geq 2 J$, the hyperbolic and the elliptic fixed 
point $s_\pm^J$ (\ref{eqn_U0gamma}) meet and annihilate themselves as illustrated in figure \ref{fig-withoutU} (f). Their disappearence is accompanied by the complete disintegration of periodic orbits. 

\begin{figure}[tb]
\centering
\includegraphics[width=5cm, angle=0]{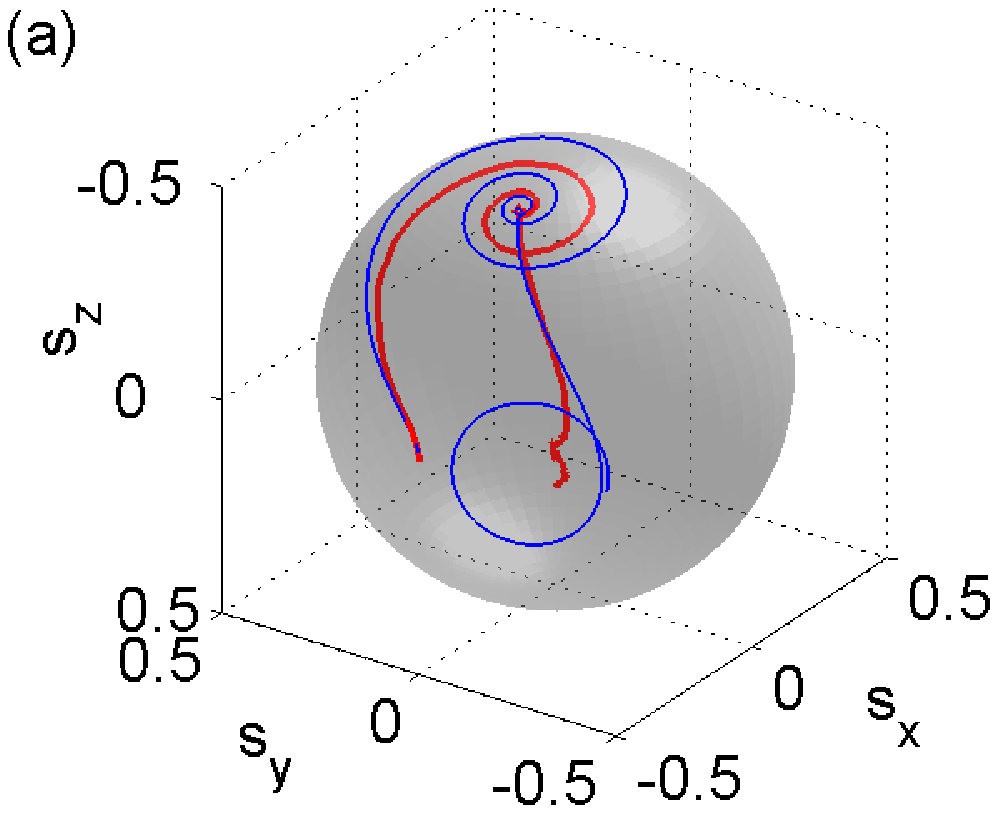}
\includegraphics[width=5cm, angle=0]{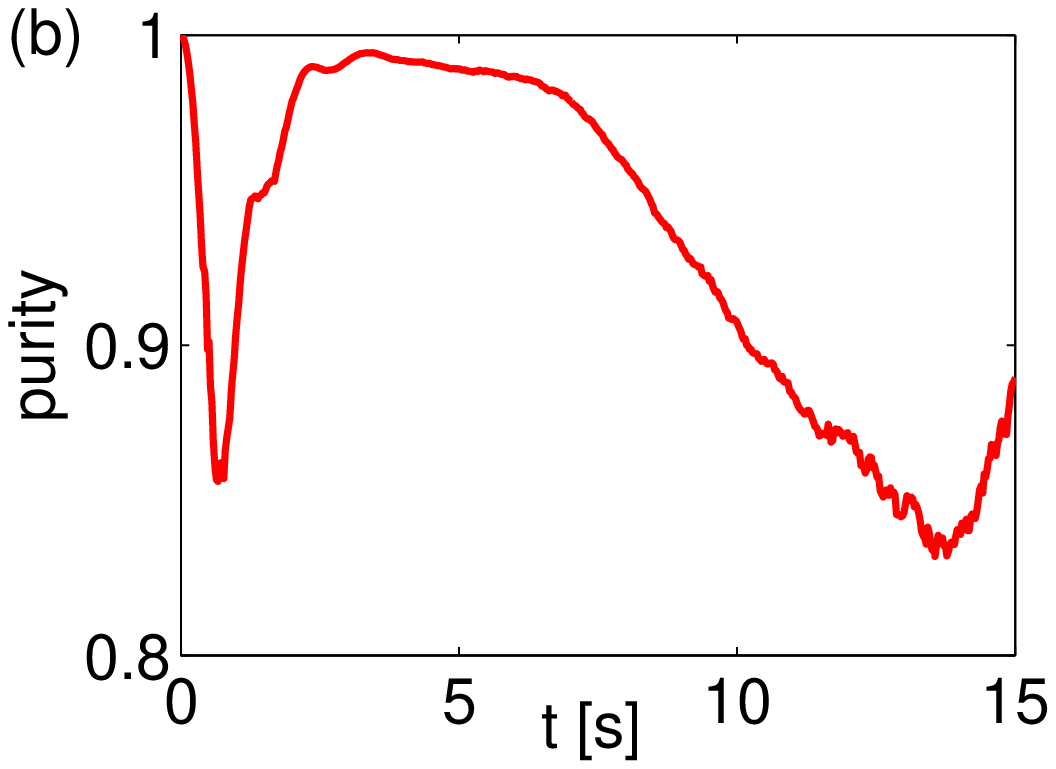}
\includegraphics[width=5cm, angle=0]{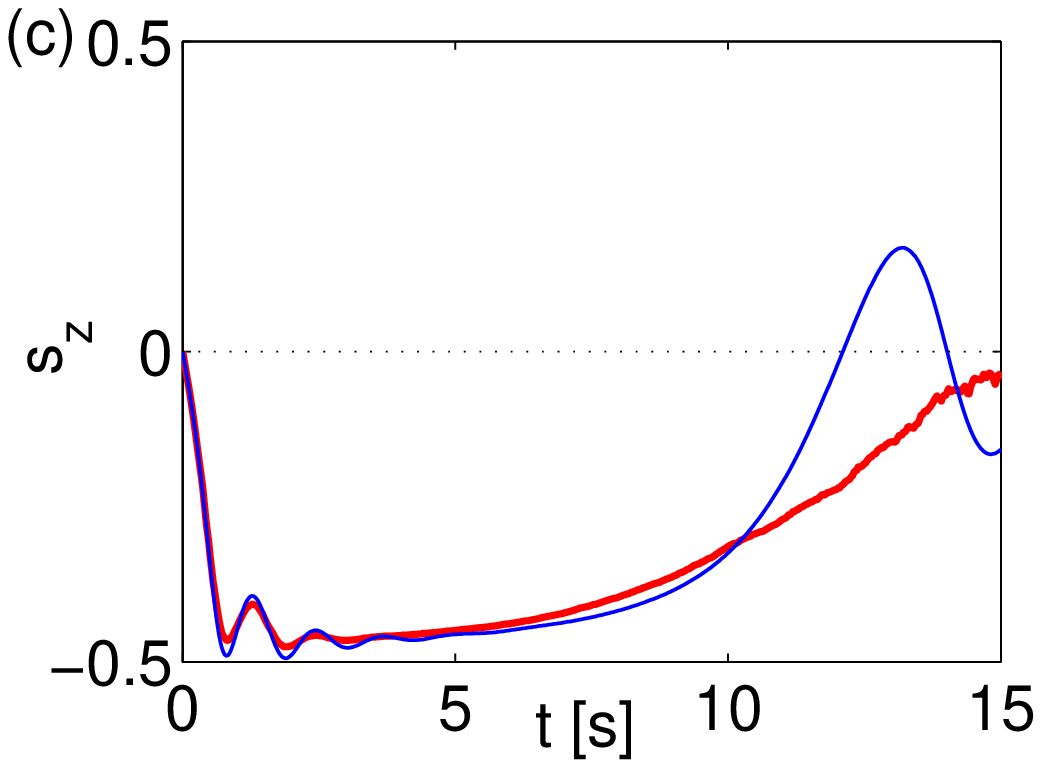}
\caption{\label{fig-reinraus}
Comparison of the many-particle dynamics (thick red line) to the
mean-field approximation (thin blues line) for an initially pure BEC.
with $\vec s = (-0.5,0,0)$ and $n(0) = 200$ particles: Dynamics
of the Bloch vector $\vec s$ (a), evolution of the purity of the BEC
(b) and evolution of the population imbalance $s_z$ (c). The system 
initially relaxes to a non-linear quasi-steady state with a purity
of almost one which is then lost as $n(t)$ decreases. Parameters 
are chosen as $J = 1 \, {\rm s}^{-1}$, $Un(0) = 10 \, {\rm s}^{-1}$, 
$T_1 = 1 \, {\rm s}$ and $f_a = 1$.}
\end{figure}

Let us finally discuss the implication of this phase space structure. We especially focus on
the emergence of the attractive fixed point since it is stable and therefore strongly 
influences the many-body quantum dynamics.
Figure \ref{fig-reinraus} shows the dynamics of the rescaled Bloch vector $\vec s$ 
comparing results of a MCWF simulation (solid red line) to the mean-field
approximation (thin blue line). The given parameters correspond to the
situations illustrated in  figure \ref{fig-withoutU} (b) or (e), respectively,
depending on the value of the macroscopic interaction strength $g(t) = Un(t)$.
The Bloch vector first relaxes to the attractive fixed point illustrated
in figure ~\ref{fig-withoutU} (e). 
The contraction of the mean-field trajectories to the attractive fixed point 
manifests itself by a convergence towards a pure BEC, which is the state of
tightest localization in phase space \cite{08phase}. This is illustrated in figure 
\ref{fig-reinraus} (b) where we have plottet the purity $\mathcal{P}:= 2 \tr (\rho^2_{\rm red})-1$ 
of the reduced single particle density matrix  $\rho_{\rm red}$, $\mathcal{P}=1$ indicating a 
pure BEC \cite{Angl01}. However, the attractive fixed point is lost as $g(t) = Un(t)$
decreases, and thus the Bloch vector departs again. This behaviour is very well
prediced by the mean-field approximation already for the modest atom number 
in the simulation.  The mean-field trajectory then convergences to the limit
cycle shown in figure \ref{fig-withoutU} (b). 
However, as the atoms are so rapidly lost nearly no particles remain to follow the 
limit cycle predicted by mean-field theory.
This transition effect between different fixed points is closely related to the 
quantum state diffusion in and out of a metastable state, which can be observed 
in optical bistability (see, e.g., \cite{Rigo97}). Note however, that the system 
considered here irretrievably departs  from the metastable self-trapping state 
because the fixed point is lost as $n(t)$ decreases.

In summary, we have derived a mean-field approximation for a dissipative 
two-mode BEC, starting from the full many-body dynamics described by a 
master equation including phase noise and particle losses. This treatment
puts the so far phenomenological description of open systems via non-hermitian 
Gross-Pitaevskii equations on a firm footing and paves the way for a variety of future 
applications in particular because the extension to an arbitrary number of modes is 
straightforward.

An analysis of the resulting equations for a fixed interaction contant $g$ shows 
that not only the critical value for the self-trapping bifucation is lowered but 
also the character of the fixed points  abruptly changes to attractive and repulsive, such 
that one of them becomes unstable. Taking into account the decline of the interaction
constant $g(t)$ due to the particle losses, the system initially converges to the 
attractive fixed point, but then suddenly jumps to a Josephson oscillation as soon as $g(t)$
falls below the critical value. This effect is understood as a manifestation 
of the metastable behaviour of the many-particle system and leads to a significant  
increase of the purity of the quantum state compared to the dissipation-free case.

The comparison to numerical results for the many-particle system obtained via the Monte 
Carlo Wave function method shows that the approach presented here is an excellent approximation 
to the full many-body dynamics already for a modest initial number of atoms. 
Thus it provides an excellent basis for a further analysis of the interplay between dissipation and 
interaction \cite{08stores}. Likewise the embedding into the more general concept 
of mean-field description \cite{Duff92} will be subject to future research.  

\ack

This work has been supported by the German Research Foundation (DFG) 
through the research fellowship programm (grant number WI 3415/1) and 
the Heidelberg Graduate School of Fundamental Physics (grant number GSC 129/1), 
as well as the Studienstiftung des deutschen Volkes.
We thank H J Korsch, E M Graefe and A Niederle for stimulating
discussions.


\end{document}